\documentclass[aps, twocolumn, showpacs]{revtex4}
\begin{document}
\title{Comment on "Spin and Orbital Angular Momentum in Gauge Theories: Nucleon
Spin Structure and Multipole Radiation Revisited"}
\author { S. C. Tiwari\\
Institute of Natural Philosophy\\
c/o 1 Kusum Kutir Mahamanapuri,Varanasi 221005, India}
\maketitle

Chen et al highlight an important issue on the nucleon spin structure
\cite{1}, and seek spin and orbital angular momentum
(SAM and OAM) operators for quarks and gluons in QCD. Since QCD as a
gauge theory is a nonabelian generalization to electromagnetism invoking QED 
paradigm is quite logical. Unfortunately the claim of resolving
the 'long standing gauge-invariance problem of the nucleon spin structure' turns out to be 
unfounded as argued below.

Some key questions \cite{2,3,4} arising in classical electrodynamics (CED) 
acquire further intricacies in QED. In CED the invariance of the action under infinitesimal
coordinate translation leads to the covariant conservation law for the canonical stress
tensor $T^{\mu \nu}$; however the AM (density) tensor $M^{\mu \nu \lambda}$ constructed
from it is not a conserved quantity. Adding a divergenceless spin energy tensor $t^{\mu \nu}$
to $T^{\mu \nu}$ a gauge invariant, symmetric, and traceless stress tensor can be obtained.
The AM tensor $A^{\mu \nu \lambda}$ constructed from symmetric tensor is conserved. 
Interestingly the AM tensor $J^{\mu \nu \lambda}$ derived as a Noether current from the
infinitesimal Lorentz rotation invariance of the action differs from $A^{\mu \nu \lambda}$
by a pure divergence term. For vanishing surface terms the equality of volume integrated AM tensor is shown, however the separation of AM into spin and orbital parts lacks gauge invariance for both.
It is well known that in QED the problem becomes more intricate: in the canonical
quantization covariant gauge condition, e. g. the Lorentz gauge $\partial _\mu A^\mu=0$
poses problems, and there arise unphysical longitudinal and timelike photons. Thus
the fundamental problem is that of satisfying two principles: manifest Lorentz covariance
and gauge invariance.The QCD AM problem is addressed generalizing QED framework; the main contribution of \cite{1}
is based on the Eqns (6) to (8). The assumption of dividing the vector potential into
physical and pure parts, and ignoring the scalar potential amounts to the loss of the covariance
at the formulation stage itself. The constraints (7) and (8), and the gauge transformation
defined by (9) and (10) restrict the gauge freedom allowed in $A_\mu ~\rightarrow ~A_\mu +
\partial _\mu \Lambda$. Recall that the radiation gauge does not pretend to solve the issue
of principles; the manipulation carried out from Eq.(6) to Eq.(11) essentially amounts to
the radiation gauge - thus the apparently gauge invariant construct (6) hardly resolves the
problem.

The SAM and OAM of light in the cited literature is based on the classical theory, and
it is misleading to invoke QED. The word photon in most of the work on AM of light is
not based on QED and has dubious physical reality \cite{5}.
Much before the advent of QED Poynting in 1909
associated AM to polarized light of magnitude $\frac {\lambda}{2\pi}$ times the linear
momentum of the wave; in the photon language it would be $\frac {\lambda}{2\pi} \frac {h\nu}
{c}=\hbar$ per photon. Beth experiment is interpreted in this sense in
ref.16 of \cite{1}, moreover the OAM of paraxial beams is based on CED and the ratio
of AM to energy using $\hbar$ is similarly interpreted OAM per photon. For the multipole radiation the fields fall off rapidly in the
radiation zone and CED predicts the ratio of the z-component of AM to energy to be $\frac {m}
{\omega}$, one says that a multipole of order $(l,m)$ carries off $m\hbar$ AM per
photon \cite{2}. Thus the light experiments cannot be construed to validate QED result, Eq.(6)
in \cite{1}.

The last point (vi) in \cite{1} is puzzling. One can write the Poynting vector as a sum of two
terms using ${\bf B}=\nabla ~\times{\bf A}$ and derive SAM and OAM \cite{6}, which correspond
to the last two terms in Eq.(6) of \cite{1}. But this does not mean that Poynting vector
does not represent momentum of the field. Though spin current is mentioned by the
authors, the subtle and most neglected question is not discussed: should spin energy
tensor $t^{\mu \nu}$ not contribute to the field energy? That it does not contribute
is eloquently discussed in \cite{3}. One consequence seems to be the ambiguity that
the energy of a photon possessing arbitrary units of OAM $(m\hbar)$ and also spin
$\hbar$ is still $\hbar \omega$. The issue of rotational energy associated with the intrinsic spin
needs attention in both QED and QCD.In conclusion, a constructive criticism is offered, 
and it is argued that the claim
of solving gauge invariance problem in \cite{1} is based on the argumentum circulo.

\end{document}